\documentclass[aps,epsfig,floats,twocolumn]{revtex4}
\usepackage{epsfig}
\begin{document}

\title{Depairing currents in superconducting films of Nb and amorphous MoGe.}

\author{A.Yu.~Rusanov, M. B. S. Hesselberth, and J.~Aarts}

\affiliation{Kamerlingh Onnes Laboratory, Leiden University, 2300 RA Leiden, The
Netherlands}

\begin{abstract}
We report on measuring the depairing current $J_{dp}$ in thin superconducting films as a
function of temperature. The main difficulties in such measurements are that heating has
to be avoided, either due to contacts, or to vortex flow. The latter is almost
unavoidable since the sample cross-section is usually larger than the superconducting
coherence length $\xi_s$ and the magnetic field penetration depth $\lambda_s$. On the
other hand, vortex flow is helpful since it homogenizes the distribution of the current
across the sample. We used a pulsed current method, which allows to overcome the
difficulties caused by dissipation and measured the depairing current in films of thin
polycrystalline Nb (low $\lambda_s$, low specific resistance $\rho$) and amorphous
Mo$_{0.7}$Ge$_{0.3}$ (high $\lambda_s$, high $\rho$), structured in the shape of bridges
of various width. The experimental values of $J_{dp}$ for different bridge dimensions are
compared with theoretical predictions by Kupriyanov and Lukichev for dirty limit
superconductors. For the smallest samples we find a very good agreement with theory, over
essentially the whole temperature interval below the superconducting critical
temperature.
\end{abstract}

\date{\today}
 \maketitle \vskip 1truecm

\section{Introduction}
The superconducting current density $J_{s}$ is a unique feature of a superconducting
material. It can be expressed as $J_{s}$=$e n_s v_s$, where $n_s$ and $v_s$ are the
density and velocity of the superconducting electrons respectively, and $e$ is the
electron charge. Increasing J$_s$ leads to increase of $v_s$ but also to a reduction of
the number of Cooper pairs. Finally, when $J_s$ reaches the depairing current $J_{dp}$,
the amount of carriers is not enough to support the supercurrent and the superconducting
state collapses. For conventional superconductors the temperature dependence of $J_{dp}$
near the critical temperature $T_{c}$ is given by the classical Ginzburg-Landau (GL)
expression $J_{dp}^{GL}(t) = J_{dp}^{GL}(0)(1-t)^{3/2}$, where $t$ = $T/T_{c}$, and
$J_{dp}^{GL}(0)$ is the depairing current at zero temperature. Early work on determining
$J_{dp}$ in Sn microbridges can be found in \cite{andrat74,skoc76}. The GL approach
becomes invalid  at lower temperatures, since the conditions $\kappa^{2}$$\gg 1-T/T_{c}$
for clean limit superconductors ($\kappa$ is Ginzburg-Landau parameter), or
$(T_{c}-T)$$\ll T_{c}$ for dirty limit superconductors, are no longer fulfilled. A more
complete and quantitative theory, valid for all temperatures and arbitrary mean free
path, was developed by Kupriyanov and Lukichev (KL), who obtained the numerical solution
of the Eilenberger equations for a superconductor carrying a current, with the velocity
of the Cooper-pairs proportional to a phase gradient of the superconducting order
parameter $\Delta$ \cite{KL80}. Notably, their theory gives the same expression for
$J_{dp}$(t) as GL theory for the temperature region close to T$_{c}$ and also yields the
correct expressions for $J_{dp}$(0) in terms of the materials constants. \\
The amount of theoretical work done on depairing currents in conventional superconductors
contrasts sharply with a lack of experimental observations. A major issue here is the
requirement with respect to sample dimensions. In principle, the sample width should not
be larger than both the penetration depth $\lambda_s$, and the coherence length $\xi_s$.
The first condition is needed to avoid current piling up at the edges, because of the
Meissner effect \cite{lik71a}. For a superconducting film $\lambda_s$ is given by
$\lambda^2_{b}/d_s$, (d$_s\ll\lambda_{b}$) where $\lambda_{b}$ is the bulk London
penetration depth, d$_s$ is film thickness, and the magnetic field is taken perpendicular
to the film plane. At low temperatures in case of dirty superconductors it becomes
$\lambda^2_{b}(\xi_{0}/\ell{d_s})$, where $\xi_{0}$ is the BCS coherence length, and
$\ell$ is the elastic mean free path. A typical value of $\lambda_{b}$ for instance for
polycrystalline Nb is 50~nm; for amorphous materials such as a-Mo$_{0.7}$Ge$_{0.3}$,
which will also be discussed below, $\lambda_{b}$ is much larger, of the order of
0.5~$\mu$m. The condition on $\xi_s$ must be fulfilled when vortex nucleation and flow is
to be prevented, which cause dissipation in sample before the $J_{dp}$ is reached. Exact
calculations made by Likharev \cite{lik71b} show that the smallest sample width below
which no vortex can appear equals 4.4$\xi_s(T)$, where $\xi_s(T)$ is the Ginzburg-Landau
coherence length given by $\xi_s(T)$ = 0.85~$\xi_s(0) / \sqrt{1-t}$, with $\xi_s(0)$ =
$\sqrt{\xi_{0}\ell}$. Typical values of $\xi_s(0)$ for our Nb and Mo$_{0.7}$Ge$_{0.3}$
are 12~nm (because of the small mean free path) and 5~nm, respectively. The only case
where both conditions can be implemented is a thin aluminum film shaped in a form of a
narrow (about 1$\mu$m) bridge. The BCS coherence length for Al is of the order of 1.5
$\mu$m, while the penetration depth can be increased to a similar value by decreasing the
film thickness. Romijn \emph{et al.} \cite{romijn82} showed that for such system the
experimental values of the depairing current density were in excellent agreement with KL
theory for temperatures down to 0.2$t$. In case of Nb and Mo$_{0.7}$Ge$_{0.3}$ films one
would have to go to a bridge width not larger than 30~nm in order to prevent vortex
appearance. \\
However, vortex motion also has an advantage, since it will homogenize the current
distribution \cite{geers01}. The main problem then in determining $J_{dp}$ is to avoid
sample heating, either by dissipation due to vortex motion, or e.g. to heating in the
contacts due to the relatively large currents. In this paper we demonstrate that the
undesired sample heating can be avoided by using a pulsed current method. We use
different superconductors, with widely different values of $J_{dp}$. Specifically,  we
use Nb with low $\lambda_b$ and also relatively low specific resistance $\rho$ (around
7~$\mu\Omega$cm) and amorphous (a-)Mo$_{0.7}$Ge$_{0.3}$ with large $\lambda_b$ and a
large $\rho \approx$ 160~$\mu\Omega$cm. Especially the large $\rho$ easily leads to
dissipation in the neighborhood of the transition to the normal metal state. Films of
different thicknesses were patterned into bridges of different width w$_s$. The
experimental values we obtain for the depairing current density $J_{dp}$(t) are in very
good agreement with the KL calculations, assuming that the current distribution across
the samples is perfectly homogeneous.

\section{Experiment}

Nb single layer films were grown by dc magnetron sputtering in an ultra high vacuum
system with a background pressure of about 10$^{-10}$ mbar and an Ar sputtering pressure
of 6$\times$10$^{-3}$ mbar. Films of a-Mo$_{0.7}$Ge$_{0.3}$ were deposited in a RF-diode
sputtering system with a background pressure of 10$^{-6}$ mbar in an Ar pressure of
8$\times$10$^{-3}$ mbar. Sputtering rates for Nb and a-Mo$_{0.7}$Ge$_{0.3}$ were 0.8
\AA/s and 1.2 \AA/s respectively. Both materials were grown on Si (100) substrates. The
thickness of the films was determined during the deposition by a crystal thickness
monitor, which was calibrated by low angle X-ray diffraction measurements and Rutherford
Backscattering. For the depairing current experiments, samples were structured in the
shape of strips of different cross-section by e-beam lithography and Ar-ion etching. The
structure included the contacts. In the case of a-Mo$_{0.7}$Ge$_{0.3}$, samples were
water-cooled during deposition and liquid nitrogen-cooled during etching, in order to
prevent undesirable film crystallization.
\begin{figure}[ht]
\begin{center}
\includegraphics[width=10cm]{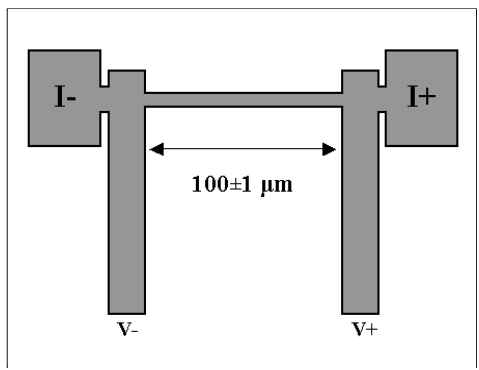}
\end{center}
\caption{Sample layout. The measurement procedure was performed with a classical 4-point
scheme. The massive current leads provide a good heat sink.}
\end{figure}
The typical geometry of the samples is shown in Fig.~1. In all cases the distance between
voltage leads was 100$\pm$1 $\mu$m. The width of resistive transition from the normal
into the superconducting state was about 30 mK for all samples. An example for both
materials is given in Fig.~2. Transport measurements in the normal state yielded an
average value of specific resistance $\rho$ of about 160 $\mu\Omega$cm for
Mo$_{0.7}$Ge$_{0.3}$ and 7.2 $\mu\Omega$cm for Nb samples respectively. For
a-Mo$_{0.7}$Ge$_{0.3}$ the elastic mean free path $\ell$ is taken to be 0.4~nm
\cite{grayb84}, of the order of the interatomic distances and these samples are clearly
in the dirty limit. For Nb, using the expressions of the free electron model with the
product $\rho \ell$ = 3.75~$\times 10^{-16}$ $\Omega$m$^2$ and the Fermi velocity $v_F$ =
5.6~$\times$10$^5$ m/s  we find $\ell$ = 5.2 nm. Comparing this value to $\xi_0$= 39 nm
for Nb \cite{weber91}, it is seen that the dirty limit condition $\ell\ll\xi_0$ is also
satisfied.
\begin{figure}[ht]
\begin{center}
\includegraphics[width=8cm]{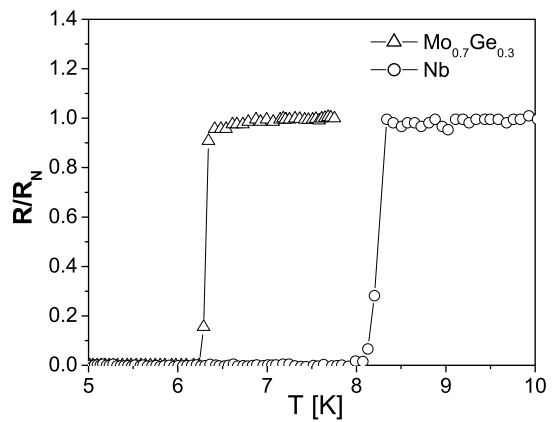}
\end{center}
\caption{Resistance normalized to its normal state value at 10 K as a function of
temperature for a Nb bridge (w$_s$ = 1 $\mu$m, d$_s$ = 20 nm) and an
a-Mo$_{0.7}$Ge$_{0.3}$ bridge (w$_s$ = 2 $\mu$m, d$_s$=64 nm) }
\end{figure}
The depairing currents measurements were performed in a $^4$He cryostat shielded from
external magnetic fields by a long permalloy (Ni$_{0.8}$Fe$_{0.2}$) screen annealed in
hydrogen atmosphere. Hall probe measurements showed a constant magnetic field background
less than $10^{-5}$ T. The samples were mounted on a massive brass holder with a
resistive heater. In order to reduce possible errors in the temperature determination
because of the temperature gradient along the sample holder, all samples were placed in
immediate proximity to the thermometer. The temperature stability during the experiment
was about 1~mK . For determination of the critical current value $I_{dp}$ at different
temperatures a pulsed current method was used, in which current pulses with a growing
amplitude were sent through the sample. The average duration of a single pulse was about
3.00$\pm$0.05~ms. Each pulse was followed by a long pause of 7.0$\pm$0.1~s. The voltage
response of the system was observed on an oscilloscope triggered for the time of a single
pulse. To improve the signal resolution a differential amplifier  was used, combined with
low-noise band filters. A typical current($I$) - voltage($V$) characteristic for
a-Mo$_{0.7}$Ge$_{0.3}$ at a reduced temperature of $t =$ 0.74 is shown in Fig.~3. One can
see a clear jump from the superconducting to the normal state at $I_{dp}$. For
temperatures close to $T_c$ a small onset of voltage was observed in all samples,
probably because of vortex motion. In order to make certain that this effect has no
influence on the determination of $I_{dp}$, the temperature was monitored during every
current pulse. Measurable differences were found very close to $I_{dp}$, as shown in
Fig.~3. We conclude that a short pulse in a combination with a long pause does not cause
sample heating and keeps the system in temperature equilibrium until the dissipation
related to the normal state occurs.
\begin{figure}[ht]
\begin{center}
\includegraphics[width=8cm]{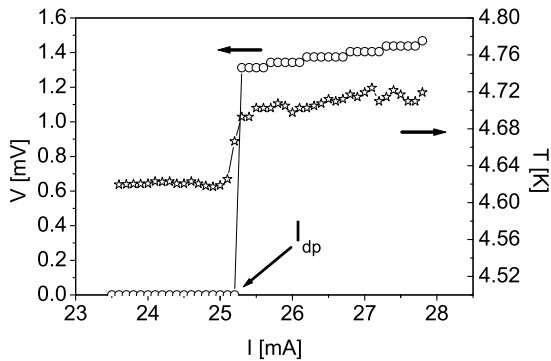}
\end{center}
\caption{Typical dependence of voltage $V$ (open circles) and temperature $T$ (open
stars) on current $I$, measured on a 2 $\mu$m wide a-Mo$_{0.7}$Ge$_{0.3}$ bridge.}
\end{figure}
\section{Results and Discussion}
To illustrate the raw data, experimentally determined values of $J_{dp}$ as a function of
reduced temperature $\textit{t}$ for two bridges of Nb (d$_s$ = 20 nm, w$_s$ = 1 $\mu$m)
and a-Mo$_{0.7}$Ge$_{0.3}$ (d$_s$ =64 nm, w$_s$ = 2 $\mu$m) are shown in Fig.~4.
\begin{figure}[ht]
\begin{center}
\includegraphics[width=7cm]{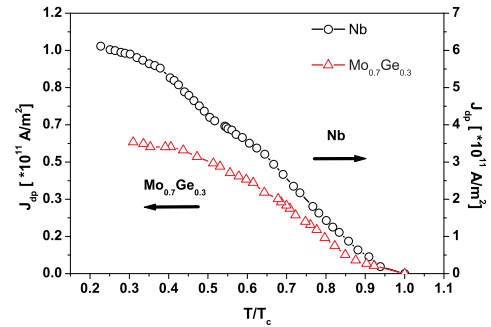}
\end{center}
\caption{Experimental results for pair-braking current $J_{dp}$ as function of reduced
temperature for a  Nb bridge (d$_s$=20 nm, w$_s$=1$\mu$m) and an a-Mo$_{0.7}$Ge$_{0.3}$
bridge (d$_s$=64 nm, w$_s$=2$\mu$m).}
\end{figure}
\begin{table}[ht]
\begin{tabular}{|c|c|c|c|c|c|c|}
\hline
Sample&d$_s$&w$_s$&$T_c$&$\rho$&$J_{dp}$(0)&J$_{dp}^{GL}$(0)\\
&[nm]&[$\mu$m]&[K]&[$\mu\Omega$*cm]&10$^{11}$[A/m$^2$]&10$^{11}$[A/m$^2$]\\
\hline
Nb&20&1.0&8.3&7.25&17&15\\
Nb&40&2.0&9.0&7.24&16&17\\
Nb&53&2.5&9.0&7.24&19&17\\
Nb&53&5.0&9.0&7.24&20&17\\
MoGe&64&2.0&6.25&160&2.0&1.6\\
MoGe&64&5.0&6.25&160&2.1&1.6\\
MoGe&64&7.0&6.25&160&2.0&1.6\\
\hline
\end{tabular}
\caption{Transport and superconducting properties of the Nb and Mo$_{0.7}$Ge$_{0.3}$
samples. Here d$_s$ and w$_s$ are the film thickness and bridge width respectively, $T_c$
is the sample critical temperature, $\rho$ is the measured specific resistance,
$J_{dp}(0)$ and $J_{dp}^{GL}(0)$ are extrapolated and calculated critical current density
at zero temperature.}
\end{table}
Between $t$ = 1 and $t$ = 0.85 both curves show a clear upturn, which indicates the
expected GL behavior. Plotting J$_{dp}^{2/3}$ as a function of $\textit{t}$ in this
temperature region results in a straight line, which can be used to extrapolate
J$_{dp}(t)$ to zero temperature. Table~1 shows the values of J$_{dp}(0)$ for all samples
investigated. It can also be used to obtain the normalized temperature dependence
(J$_{dp}(t)$/$J_{dp}$(0))$^{2/3}$, which has a universal form in KL theory. Plots of this
quantity for samples with different bridge width are shown in Fig.~5 for Nb and in Fig.~6
for a-Mo$_{0.7}$Ge$_{0.3}$. Both the absolute values of J$_{dp}(0)$ and the temperature
dependence can be directly compared to the KL results, which we now briefly
reiterate. \\
Close to $T_c$ the depairing current density can be written as follows :
\begin{center}
\begin{equation}
J_{dp}^{GL}(t) =1.93{\chi^{1/2}(\rho)}eN(0)\upsilon_Fk_BT_c(1-T/T_c)^{3/2}
\end{equation}
\end{center}
where $\chi(\rho)$ is the Gor'kov function controlled by a dimensionless parameter
characterizing the amount of electron scattering, $\rho$=($\hbar\upsilon_F$)/(2$\pi
k_BT_c\ell$), with $\ell$ the elastic mean free path and N(0) the density of states at
the Fermi level for each spin direction. For $\ell\ll\xi_0$ (dirty limit)
$\rho\rightarrow\infty$, which yields for $\chi(\rho)\rightarrow 1.33\ell/\xi_0$. Thus,
at zero temperature the extrapolated depairing current density $J_{dp}^{GL}$(0) becomes
\begin{center}
\begin{equation}
J_{dp}^{GL}(0) = 1.26eN(0)\upsilon_F\Delta(0)\sqrt{\frac{\ell}{\xi_0}} \label{eq:Jdp0}
\end{equation}
\end{center}
Because of the small mean free path in both types of samples, we may assume applicability
of the free-electron model, so the density of states N(0) can be expressed as
\begin{center}
\begin{equation}
N(0) = (\frac{2}{3}e^2\upsilon_F\rho\ell)^{-1}
\end{equation}
\end{center}
Substituting this formula in Eq.~\ref{eq:Jdp0} with
$\xi_0=\hbar\upsilon_F$/$\pi\Delta(0)$ and $\Delta(0)=1.76k_BT_c$ we obtain
\begin{center}
\begin{equation}
J_{dp}^{GL}(0) = 244\biggl[\frac{(T_c)^3}{\upsilon_F(\rho\ell)\rho}\biggr]^{1/2}
\label{eq:JdpGL-final}
\end{equation}
\end{center}
This result is similar to the one obtained in \cite{romijn82} and \cite{geers01}.
Eq.~\ref{eq:JdpGL-final} contains only experimental quantities and the $\rho\ell$
product, which is known for both materials from literature
\cite{grayb84,weber91,grayb87}.
\begin{figure}[ht]
\begin{center}
\includegraphics[width=7cm]{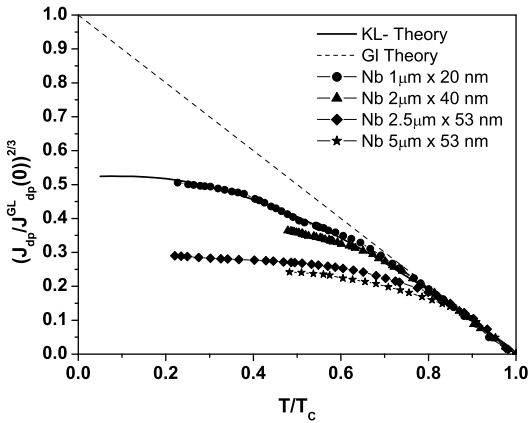}
\end{center}
\caption{Experimental results for the pair-braking current density $J_{dp}$ normalized to
its extrapolated value $J_{dp}(0)$ as a function of reduced temperature in Nb bridges of
different width and thickness as denoted. The black solid and dashed lines indicate KL
and GL results respectively.}
\end{figure}
Looking now at Figs.~5 and 6, all curves follow GL behavior down to about $t$~= 0.85. The
values of J$_{dp}(0)$ extrapolated from this region can be compared to the values
calculated from Eq.~\ref{eq:JdpGL-final}  for $J_{dp}^{GL}$(0). This comparison is made
in Table~1 which gives all relevant parameters for the different samples. Basically, we
find quite good agreement for all sample widths. In the case of Nb, the most serious
deviation is found for the 5~$\mu$m bridge, which is presumably due to contact heating as
a result of the larger current. It is interesting to note that the extrapolated values
are the same as found by Geers {\it et al.} \cite{geers01} who used continuous currents
and larger bridge widths. The differences are in the extent of the GL-regime, which was
only found down to $t$~= 0.93 in the earlier experiments, and also in the temperature
dependence below the GL regime. There, the temperature dependence is described by the
full KL-calculation, which was also performed in ref.~\cite{geers01}. For a single
superconducting film, the results for Nb are shown in Fig.~5 by the solid line. The
smallest sample (d=20 nm, w=1~$\mu$m) follows the KL theoretical curve down to $t$ = 0.2
without a significant deviations. Wider bridges show a suppression of $J_{dp}(t)$ with
respect to the calculated value, again in agreement with earlier results \cite{geers01}.
Presumably, sample heating via contacts and vortex flow occurs even for the short time of
a current pulse. It appears therefore that using low (pulsed) currents, $J_{dp}(t)$ can
be determined correctly over the full temperature range for other materials than Al.
\begin{figure}[ht]
\begin{center}
\includegraphics[width=7cm]{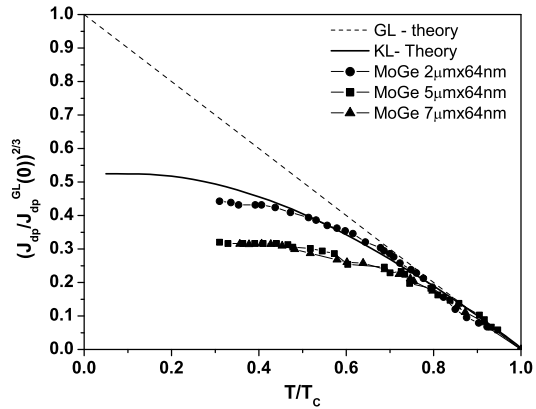}
\end{center}
\caption{Experimental results for the pair-braking current density $J_{dp}$ normalized to
extrapolated value $J_{dp}(0)$ as a function of reduced temperature in
Mo$_{0.7}$Ge$_{0.3}$ bridges of different width and thickness as denoted. The black solid
and dashed lines indicate KL and GL results respectively.}
\end{figure}
Circumstances can be somewhat less favorable, however, as shown by the measurements on
a-Mo$_{0.7}$Ge$_{0.3}$. These were performed only for a film thickness of 64 nm. In the
GL regime the difference between measured and calculated values of $J_{dp}(0)$ is
somewhat larger than for Nb (see Table 1), with the measured values larger than the
calculated ones. It will be clear that this cannot be due to pile-up of current at the
samples edges, which would yield the opposite effect. Moreover, for amorphous materials
this should be less of a problem, since the penetration depths are very large and
actually of the order of the smallest bridge width. The difficulty rather lies in the
correct determination of $J_{dp}(t)$ close to $T_c$, with more scatter in the individual
points. One reason for this may be the very low vortex pinning which is characteristic of
amorphous materials \cite{grayb86,berg93}. Another may be that the processing of the film
during the structuring process may lead to changes in the material. For instance, the
specific resistance we find for the bridges is about 10~\% lower than for wider
structures \cite{plourde02}. Also, thinner films showed increasing $\rho$ and decreasing
$T_c$, which in this thickness regime cannot be well explained by the onset of
localization effects \cite{grayb84}. Since amorphous materials are very sensitive to
recrystallization, this may be playing a role. Still, the difference between $J_{dp}(0)$
and $J_{dp}^{GL}(0)$ is only 20~\%, which may still be considered very good. For the
temperature dependence (Fig.~6) the result is also similar to Nb. For the smallest
bridge, the experimental curve shows good agreement with the theoretical prediction,
while for wider bridges the values remain too low. \\
\newline
In summary, we have shown that measurements of depairing currents in  conventional
type-II superconductors with cross-section larger than their characteristic lengths
$\xi_s$ and $\lambda_s$ is well possible by using a pulsed current method. Using two
different superconductors with quite different values of their depairing current, we
found good agreement between experiments and theory with respect to both the absolute
values and the temperature dependence, over essentially the full range of temperatures.
Such an unambiguous determination of a quantity which directly measures the
superconducting order parameter should also find use in problems posed by hybrid systems;
in particular, it will be the correct quantity to gauge the effects of suppression of
superconductivity by the injection of (spin-polarized) quasiparticles.

\section{Acknowledgement}
This work is part of the research program of the "Stichting voor Fundamenteel Onderzoek
der Materie (FOM)", which is financially supported by NWO. We would like to thank A. A.
Golubov for his calculation of the depairing current,  V.V. Ryazanov and R. Besseling for
helpful discussions, and S. Habraken for assistance in the experiments.

\end{document}